\documentclass{elsart}

\usepackage{natbib}

\usepackage{graphics}
\usepackage{graphicx}
\usepackage{epsfig}

\usepackage{amssymb}
\usepackage{latexsym}

\begin{document}

\begin{frontmatter}



\title{The WSRT Virgo Filament Survey}


\author{A. Popping}
\ead{popping@astro.rug.nl}
\address{Kapteyn Astronomical Institute, P.O Box 800, 9700 AV Groningen, The Netherlands}
\author{R. Braun}
\address{ASTRON, P.O. box 2, 7990 AA Dwingeloo, The Netherlands}

\begin{abstract}
In the last few years the realization has emerged that the universal
baryons are almost equally distributed by mass in three components:
(1) galactic concentrations, (2) a warm-hot intergalactic medium
(WHIM) and (3) a diffuse intergalactic medium. These three components
are predicted by hydrodynamical simulations and are probed by QSO
absorption lines. To observe the WHIM in neutral hydrogen,
observations are needed which are deeper than log(N$_{HI}$)=18. The
WHIM should appear as a Cosmic Web, underlying the galaxies with
higher column densities. We have used the WSRT, to simulate a filled
aperture by observing at very high hour angles, to reach very high
column density sensitivity. To achieve even higher image fidelity, an
accurate model of the WSRT primary beam was developed. This will be
used in the joint deconvolution of the observations. To get a good
overview of the distribution and kinematics of the Cosmic Web, a deep
survey of 1500 square degrees of sky was undertaken, containing the
galaxy filament extending between the Local Group and the Virgo
Cluster. The auto-correlation data has been reduced and has an RMS of
$\Delta N_{HI} = 4.2\times10^{16}$ cm$^{-2}$ over
20~kms$^{-1}$. Several sources have been tentatively detected, which
were previously unknown, as well as an indication for diffuse
intergalactic filaments.
\end{abstract} 


\end{frontmatter}

\section{Introduction}
\label{}
The number of detected baryons in the Low Redshift Universe is
significantly below expectations. According to cosmological
measurements the baryon fraction is about 4$\%$ at $z \sim 2$
(\cite{2003ApJS..148....1B}; \cite{2003ApJS..148..175S}). This is
consistent with actual numbers of baryons detected at $z > 2$
(\cite{1997ApJ...490..564W}; \cite{1998ARA&A..36..267R}). In the
current epoch however, at $z \sim 0$ about half of this matter has not
been observed (\cite{1998ApJ...503..518F}; \cite{1999ApJ...514....1C};
\cite{2000ApJ...534L...1T}; \cite{2002ApJ...564..631S};
\cite{2004ApJS..152...29P}).

Recent hydrodynamical simulations give a possible solution for the
¨Missing Baryon¨ problem (\cite{1999ApJ...514....1C};
\cite{2001ApJ...552..473D}; \cite{2002ApJ...564..604F}). These
simulations show that in the current epoch, cosmic baryons are almost
equally distributed amongst three phases \citep{1999ApJ...511..521D}:
(1) diffuse phase, (2) a shocked phase and (3) a condensed phase. The
diffuse phase is associated with warm, low-density photo-ionized
gas. The shocked phase consists of gas that has been heated by shocks
during structure formation, with a moderate overdensity. Because the
temperature range is very broad from $10^5$ to $10^7$ K, it is also
called the Warm Hot Intergalactic Medium (WHIM). The condensed phase
is associated with cool galactic concentrations. These three components are
each coupled to a decreasing range of baryonic over-density:
$log(\rho_H/\bar{\rho}_H)$ $<$ 1, 1-3.5, and $>$ 3.5 and are probed by QSO
absorption lines with specific ranges of neutral column density:
log$(N_{HI})$ $<$ 14, 14-18 and $>$ 18 \citep{2005ASPC..331..121B}.

\section{WHIM}
The Warm Hot Intergalactic Medium is thought to be formed during
structure formation. Low density gas is heated by shocks 
during its infall onto the filaments that define the Large Scale
Structure of the Universe. Most of these baryons are still
concentrated in unvirialized filamentary structures of highly
ionized gas.

The WHIM has been observationally detected in O VI absorption
\citep{astro-ph/0411151}, but also via Ne VIII
\citep{2005ApJ...626..776S} and X-ray absorption
\citep{2005AdSpR..36..721N}. Of course absorption studies alone, do
not give us complete information on the spatial distribution of the
WHIM. Emission from the WHIM would give entirely new information about
the distribution and kinematics.

Direct detection of the WHIM is very difficult in the EUV and X-ray
bands \citep{1999ApJ...514....1C}. The gas is ionized to such a
degree, that it becomes ``invisible'' in infrared, optical or UV light,
but should be visible in the FUV and X-ray bands
\citep{2005AdSpR..36..721N}. Given the very low density, extremely
high sensitivity and a large field of view is needed to image the
filaments. Capable detectors are not yet available for the X-ray or
FUV (\cite{2003PASJ...55..879Y}; \cite{2005AdSpR..36..721N}).

Large-scale gas filaments have been detected in X-ray emission
(\cite{1997ApJ...487L..13W}; \cite{2000ApJ...528L..73S};
\cite{2001ApJ...563..673T}). With the current facilities, these X-ray
emission studies reveal gas which is hotter and denser than the
WHIM. This gas is not expected to contain a substantial part of WHIM
baryons at the current epoch \citep{2001ApJ...552..473D}. The WHIM should be most prominent in the soft X-ray
band. The interstellar medium of the Galaxy complicates its detection
since in the soft X-ray band the Galaxy is a strong source of emission
at an effective temperature of $\sim10^6$ K. The neutral hydrogen in
the Galaxy on the other hand, prevents observations of the EUV.

We have adopted a different approach by observing the WHIM in neutral
hydrogen emission.

At the current epoch we can confidently predict that in going down
from HI column densities of $10^{19}$ cm$^{-2}$ (which define the
current ¨edges¨ of well studied nearby galaxies in HI emission) to
$10^{17}$ cm $^{-2}$ the surface area will increase by a factor of 30
\citep{2004A&A...417..421B}. The critical observational challenge is
crossing the ``HI desert'', the range of log(N$_{HI}$) from about 19.5
down to 18 over which photo-ionization by the intergalactic radiation
field produces an exponential decline in the neutral fraction from
essentially unity down to a few percent
(eg. \cite{1994ApJ...423..196D}). Nature is kinder again to the HI
observer below log(N$_{HI}$) = 18, where the neutral fraction
decreases only very slowly with log(N$_{HI}$). The neutral fraction of
hydrogen is thought to decrease with decreasing column density from
about 100$\%$ for $log(N_{HI})$ $\>=$ 19.5 to about 1$\%$ at
$log(N_{HI})=17$ \citep{1994ApJ...423..196D}. The baryonic mass traced
by this gas (with a 1$\%$ or less neutral fraction ) is expected to be
comparable to that within the galaxies, as noted above.

\begin{figure}
  \includegraphics[width=1.0\textwidth]{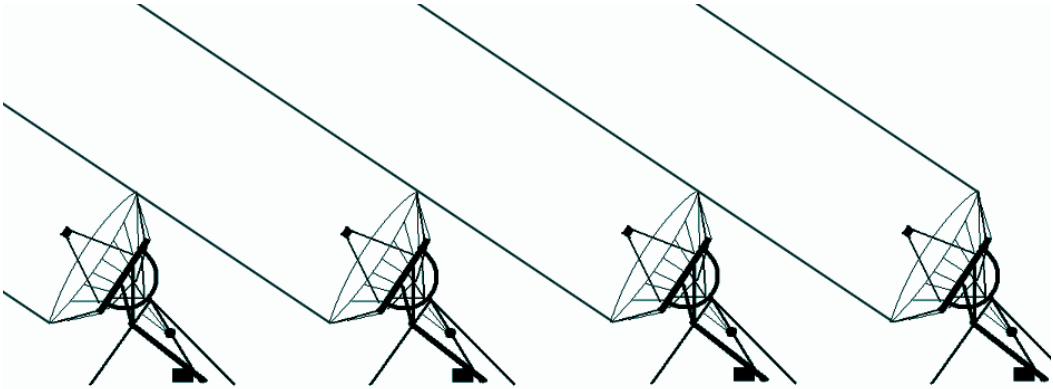}
  \caption{Observing mode of the WSRT dishes, a filled aperture of 300
  m can be simulated when observing at large hour angles.}
  \label{config}
\end{figure}

\section{Observations}
To obtain the highest possible brightness sensitivity, the WSRT has
been used to simulate a large filled aperture. 12 of the 14 WSRT 25~m
telescopes are positioned at regular intervals of 144~m. When
observing at very low declinations and extreme hour angles, a filled
aperture can be formed (as can be seen in Fig.~\ref{config}), which is
300 $\times$ 25 m in projection. In this peculiar observing mode the
excellent spectral baseline and PSF properties of the interferometer
are still obtained while achieving excellent brightness sensitivity. A
deep fully-sampled survey of the galaxy filament joining the Local
Group to the Virgo Cluster has been undertaken, extending from 8 to 17
hours in RA and from -1 to +10° in Dec. and covering 40~MHz of
bandwidth with 8~km~s$^{-1}$ resolution. Mosaic-mode drift-scan
observations are undertaken twice for each of 22,000 positions, once
at positive hour angle and once at negative hour angle. The expected
survey RMS sensitivity is $\Delta N_{HI} \sim 2\times10^{17}$
cm$^{-2}$ over 20 km~s$^{-1}$ in the synthesis data.

\begin{figure}
  \includegraphics[width=1.0\textwidth]{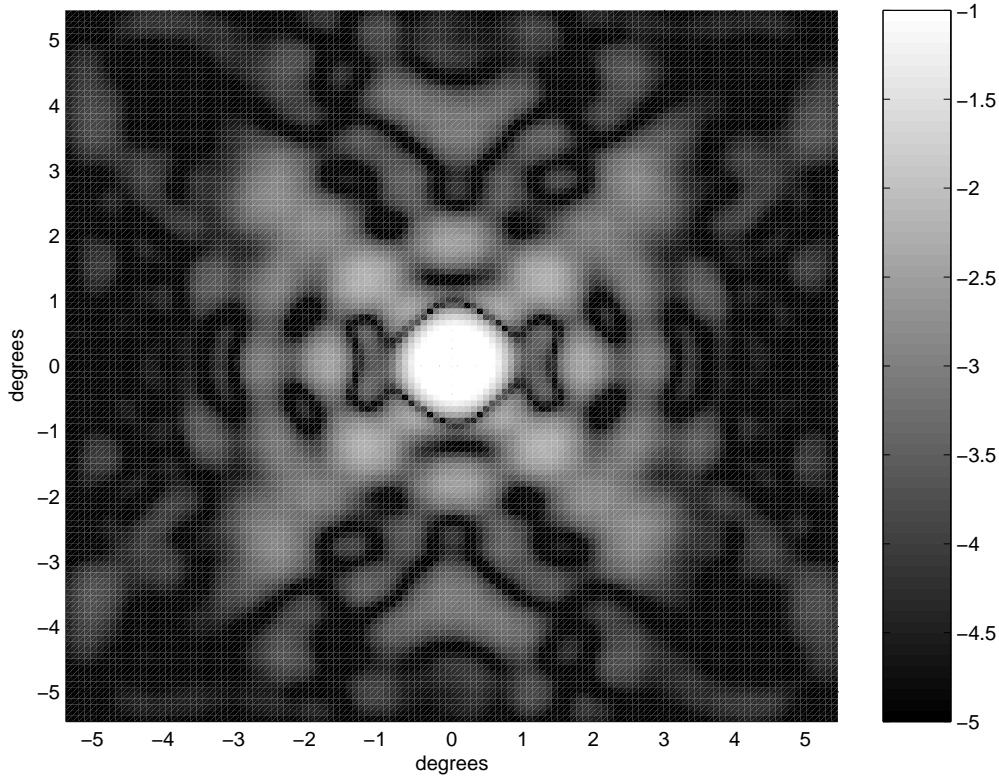}
  \caption{The Point Spread Function (PSF) of the primary beam of the
  WSRT on logarithmic scale, averaged over  a full band.}
  \label{beampsf}
\end{figure}

\subsection{Primary Beam}
\begin{figure}
  \includegraphics[width=0.5\textwidth]{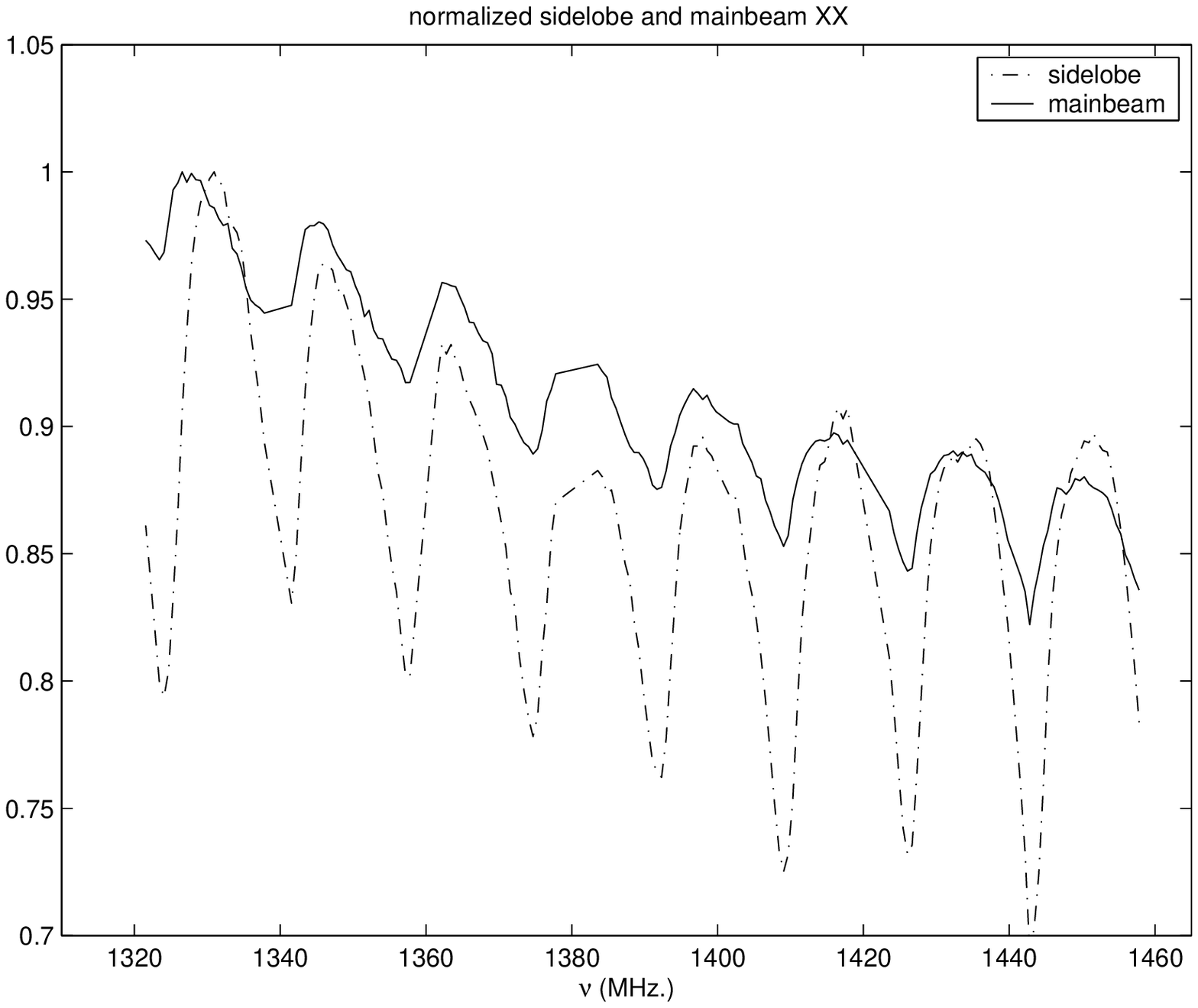}
  \includegraphics[width=0.5\textwidth]{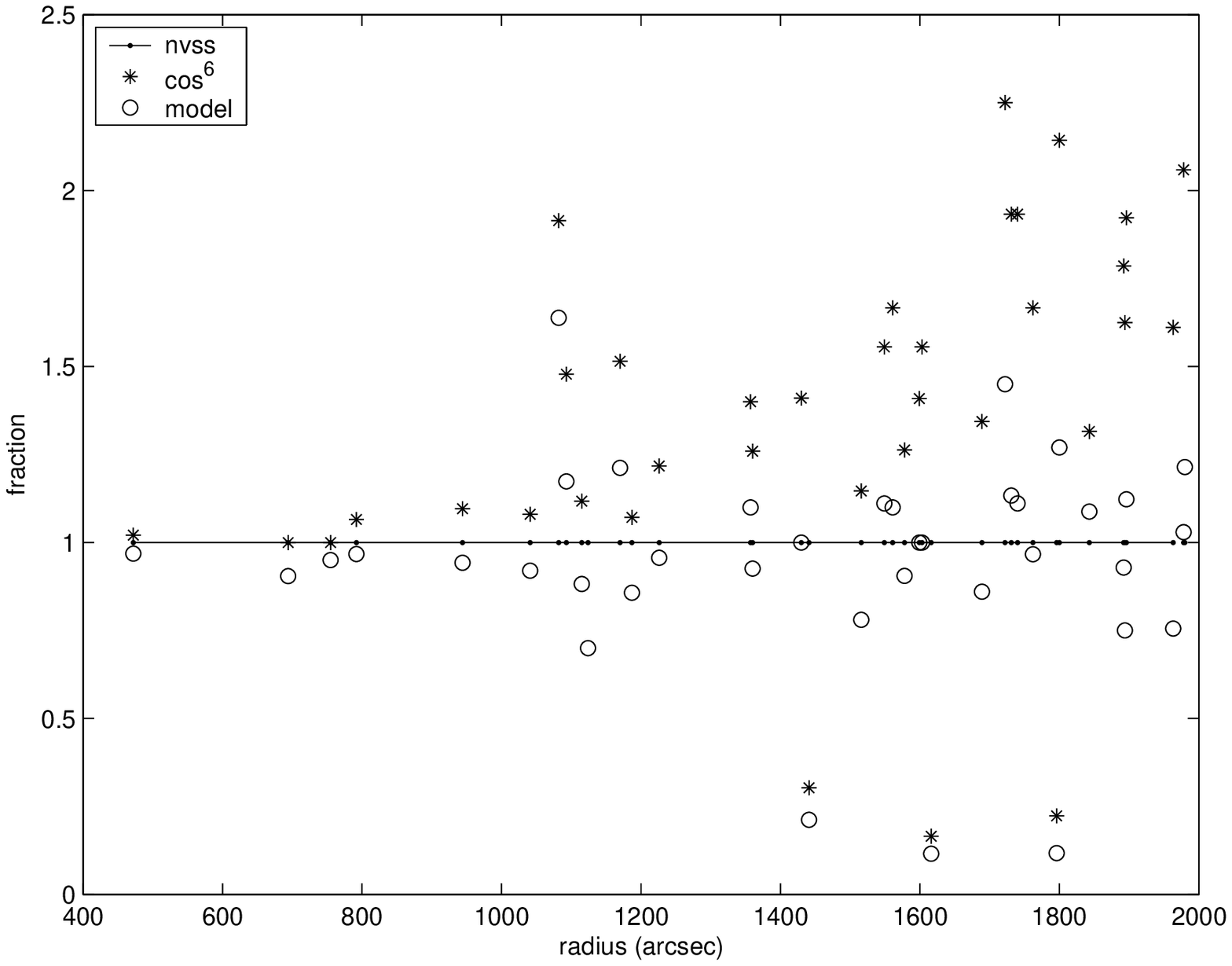}
  \caption{Left: The $\sim$17 MHz period for the integrated power of
  the main lobe and the side lobes. Right: The normalized NVSS fluxes
  with the old and new beam descriptions, especially at large radii
  from the beam center, the new model gives a significantly better
  correction for the Primary Beam Attenuation}
  \label{beam}
\end{figure}

To obtain high dynamic range and image fidelity in a mosaiced field,
the shape of the primary beam must be known to high precision. The
approximation for the beam shape used in the past has strong
limitations, since it assumes the beam to be axisymmetric and is
truncated at a level of a few percent, which is above the level of the
first sidelobes. We modeled the WSRT primary beam in great
detail. Three series of holographic measurements were undertaken to
fully sample angular scales between a few arcminutes and several
degrees. An empirical model-cube was constructed, containing the
primary beam pattern between 1322 and 1457 MHz, with a 1 MHz frequency
resolution. In Fig.~\ref{beampsf} the model is plotted on a
logarithmic scale, averaged between 1424 and 1440 MHz. In Fig.~\ref{beam} (left panel) the
integrated main beam and side lobes are plotted on a normalized scale
against frequency. There is a very significant periodic modulation of
the beam size of $\sim 4\%$ in addition to the
expected linear scaling with frequency. This modulation has a period
of $\sim$17 MHz. When doing the joint-deconvolution of the mosaic
observations, an interpolated PSF will be calculated, corresponding to
the frequency of the data.

Compared to the previously employed cos$^6$ function, the new model
has a systematically broader main beam (at all phases of the
oscillations), a ``diamond''-shaped departure from circular symmetry as
well as inclusion of the first few near-in sidelobes (and their strong
variation with frequency). When correcting for the primary beam
attenuation, this can have a significant influence on the calculated
source fluxes, especially far from the beam center. To investigate the
difference between the new model and the old beam description we have
compared the implied flux densities of off-axis continuum sources
with those obtained from the NVSS survey.

The NVSS (NRAO VLA Sky Survey: \cite{1998AJ....115.1693C}) is a large
area radio survey at 1.4 GHz. The images are on-line and can be
downloaded. We have employed the WSRT SINGS continuum survey \citep{braun2006} which contains single pointing observations of more than 30
fields at the same frequency which means that they can be easily
compared. We have produced an inital sample of 36 compact sources for
which fluxes have been determined in both the NVSS images as well as
the WSRT single pointing observations. The WSRT observations have been
corrected for primary beam attenuation with the old cos$^6$
approximation and with the new model. The sources are located at
different distances from the pointing center, to determine the
reconstructed flux accuracy at different radii.

The results are plotted in Fig.~\ref{beam} (right panel), where the
obtained fluxes are plotted against the radius from the center of the
beam. All fluxes are normalized to the fluxes of the NVSS observations
which are used as a reference. At small radii, the old approximation
(stars) as well as the new model (open circles) are very comparable
with the NVSS-fluxes (straight line). When going to larger radii,
there are more differences. The main beam of the cos$^6$ approximation
is clearly too small, which means that the correction for attenuation
will be larger and flux discrepancies will be larger when going out
from the center. Indeed, at large radii the old approximation gives fluxes
which are significantly too high, while the new model agrees
with the reference values. Much larger samples of comparison sources
will be required to document the accuracy of the new correction in detail.

\section{Results}
\begin{figure*}
  \includegraphics[angle=-90, width=1.0\textwidth]{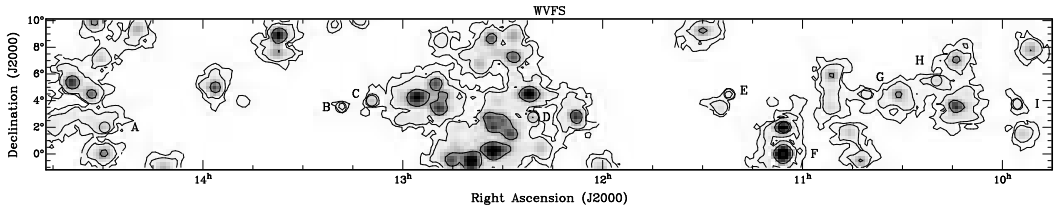}
  \caption{Column density map with contour levels at $8\times10^{17}$, $3\times10^{18}$ and $3\times10^{19}$ cm$^{-2}$. The new objects are encircled and assigned a letter. }
  \label{wvfs}
\end{figure*}

All observations for our WSRT Virgo Filament Survey (WVFS) have now
been completed. We have begun analysis of the database by first
reducing the auto-correlations which were obtained simultaneously with the
cross-correlation mosaic. The first data cubes and moment images are
now available for this total power data. Figure ~\ref{wvfs} shows the
integrated column density map of the observed region. In addition to
the many hundreds of known galaxies which were detected in this region
there are a total of nine discrete sources found, which have not been detected before. All sources were selected to be above a level of $7\sigma$. The radial velocities are corrected for Virgo infall and the distances are calculated, using $H=71$ km s$^{-1}$ Mpc$^{-1}$. Global properties for the nine objects are given in table~\ref{sources}.\\
Between the galaxies, diffuse filaments are apparent with column
densities going down to $\sim8 \times 10^{17}$ cm$^{-2}$. While this
is exactly the type of signature we were hoping to detect, further
analysis will be required to rule out an instrumental origin for these
features.

\begin{table*}
\begin{tabular}{|c|ccccccc|}
\hline
 & RA & DEC & Flux & max(N$_{HI}$) & Vrad & Distance & Mass\\
 &  &  & (Jy-km/s) & (cm$^{-2}$)  & km/s &(Mpc) & $M_{\odot}$\\

\hline
A & 14:29:19 & 02:00:00 & 26 & 1.8$\times10^{19}$ & 1604 & 22.6 & 3$\times10^{9}$\\
B & 13:18:08 & 03:30:00 & 4  & 5$\times10^{18}$  & 791  & 11.1 & 1$\times10^{8}$\\
C & 13:09:06 & 04:00:00 & 18 & 2$\times10^{19}$ & 938  & 13.2 & 7$\times10^{8}$\\
D & 12:20:59 & 02:45:00 & 22 & 3$\times10^{19}$  & 1350 & 19.0 & 2$\times10^{9}$\\
E & 11:21:49 & 04:30:00 & 5  & 5$\times10^{18}$  & 1389 & 19.6 & 4$\times10^{8}$\\
F & 11:05:46 & 02:00:00 & 131& 1.4$\times10^{20}$ & 785  & 11.0 & 4$\times10^{9}$\\
G & 10:40:42 & 04:30:00 & 3 & 5$\times10^{18}$ & 596 & 8.4 & 6$\times10^{7}$\\
H & 10:19:38 & 05:30:00 & 6  & 9$\times10^{18}$  & 1167 & 16.4 & 4$\times10^{8}$\\
I & 09:55:35 & 03:45:00 & 8  & 9$\times10^{18}$  & 1344 & 18.9 & 7$\times10^{8}$\\
\hline
\end{tabular}
\caption{Global properties of the detected sources.}
\label{sources}
\end{table*}

\section{Discussion}
The Warm Hot Intergalactic medium is expected to contribute about
30-50~$\%$ to the total baryon budget. Observing the WHIM is very
difficult and extremely sensitive methods are needed. We are observing
the WHIM in neutral hydrogen emission since the expected column densities,
between log(N$_{HI}$) 14-18, are just accessible to current instrumentation at the high end. The first results look promising, since
we reach a sensitivity of $\Delta N_{HI} = 4.2 \times 10^{16}$
cm$^{-2}$ over 20 km s$^{-1}$ in the auto-correlation data of our survey.\\
Nine discrete sources have been tentatively detected which were previously unknown, as well as an indication for diffuse filaments that may be the HI counterparts of the "Cosmic Web".\\
Once the distribution of the WHIM is known, absorption lines
toward strong background sources will be sought. These may provide
both the metalicity and the neutral fraction. With the neutral
fraction in hand, the total mass of the WHIM can be estimated, allowing us to test the conjecture that the "Missing Baryons" have finally been found.








\bibliographystyle{aa}
\bibliography{names,wvfsbibliography}

\end{document}